\documentclass[preprint, letter]{aastex}
\usepackage{graphicx}
\usepackage{subfigure}
\usepackage{amsmath}
\usepackage{color}
\usepackage{natbib}

\newcommand{\mytable}[1]{Table \ref{#1}}

\newcommand{\mycitet}[1]{\citet{#1}}
\newcommand{\mycitep}[1]{\citep{#1}}
\newcommand{\myeq}[1]{Eq. \ref{#1}}
\newcommand{\myfig}[1]{Fig. \ref{#1}}

\newcommand{\intd}[1]{\ensuremath{\,\mathrm{d}#1}}
\newcommand{\dd}[2]{\frac{\partial #1}{\partial #2}}

\title{Design of mirrors and apodization functions in phase-induced amplitude apodization (PIAA) systems}
\author{Eric Cady}
\affil{Jet Propulsion Laboratory, California Institute of Technology\footnote{Copyright 2012 California Institute of Technology. Government sponsorship acknowledged.}}
\affil{4800 Oak Grove Drive, Pasadena, CA, 91109}
\email{eric.j.cady@jpl.nasa.gov}

\begin{abstract}
Phase-induced amplitude apodization (PIAA) coronagraphs are a promising technology for imaging exoplanets, with the potential to detect Earth-like planets around Sun-like stars.  A PIAA system nominally consists of a pair of mirrors which reshape incident light without attenuation, coupled with one or more apodizers to mitigate diffraction effects or provide additional beam-shaping to produce a desired output profile.  We present a set of equations that allow apodizers to be chosen for any given pair of mirrors, or conversely mirror shapes chosen for given apodizers, to produce an arbitrary amplitude profile at the output of the system.  We show how classical PIAA systems may be designed by this method, and present the design of a novel 4-mirror system with higher throughput than a standard 2-mirror system.  We also discuss the limitations due to diffraction and the design steps that may be taken to mitigate them.
\end{abstract}


\begin{document}
\maketitle

\section{Introduction}

The challenge of imaging an Earth-like planet around a Sun-like star has two primary components: contrast and angular resolution.  An Earth twin orbiting a sun twin at 1 AU at a distance of 10 parsecs will be $10^{10}$ times fainter than its host star, while being separated by only 100 milliarcseconds.  For a 2m telescope with a circular aperture operating at 500nm---a reasonable expectation for a visible-band space telescope---this places the planet at $1.95 \lambda/D$ in the image plane.  Here, the intensity in the point spread function (PSF) from the host star will be $\sim6\times10^{-3}$ lower than the peak of the diffraction pattern and over six orders of magnitude higher than the planet, making detection of the planet virtually impossible.  To overcome this, a number of optical systems---including coronagraphs, occulters, and interferometers---have been proposed to reduce the residual stellar light in the regions of the image plane where we wish to look for planets.

One of the most promising coronagraphs for space applications is the phase-induced amplitude apodization (PIAA) coronagraph \mycitep{Guy03, Tra03, Guy05, Van05}, which uses a pair of mirrors to losslessly remap incident light to apodize the beam while retaining high throughput.  In addition, incident wavefronts with an angle to the optical axis of the system are displaced disproportionately from the optical axis, providing an effective magnification of the inner working angle.

A number of these systems have been designed \mycitep{Plu06} and tested \mycitep{Bel09, Ker09, Guy10, Bal10, Pue11a}, in conjunction with apodizing elements to reduce manufacturing requirements on mirrors or mitigate diffraction from mirror edges.  As shown in \mycitet{Van06}, without at minimum an apodizer before the mirrors (a \emph{pre-apodizer}), the effects of diffraction from the edges of the mirrors limit the achievable contrast severely, to $10^{-5}$.  Apodizers after the mirrors (a \emph{post-apodizer}) have also proved useful for performing a portion of the beam-shaping, at the expense of some throughput; using a post-apodizer can reduce the manufacturing requirements on the mirrors, particularly the curvatures required on the edges of some designs \mycitep{Plu06}.

While the tools for accurately modeling propagation through PIAA systems have become fairly sophisticated and robust \mycitep{Bel06a, Pue09, Kri10, Pue11}, the tools for finding new apodizations and mirror shapes are not as developed; in particular, the interaction between multiple remapping and apodizing components in the design has not been fully explored.  In this paper, we discuss the propagation of radially-symmetric amplitude functions through PIAA systems, in particular setting down a series of equations in Sec. \ref{sec:3f} which allow for the arbitrary specification of amplitude profiles throughout the system.  This permits mirrors and apodizers to be designed explicitly to work jointly as a system; examples of this are shown in Sec. \ref{sec:app}.  Applications to design and modeling of four-mirror systems are discussed in Sec. \ref{sec:4m}, including their interactions with apodizers in the system.

\section{Design of amplitude profiles} \label{sec:3f}

\begin{figure}
\begin{center}
\includegraphics[width=6.5in]{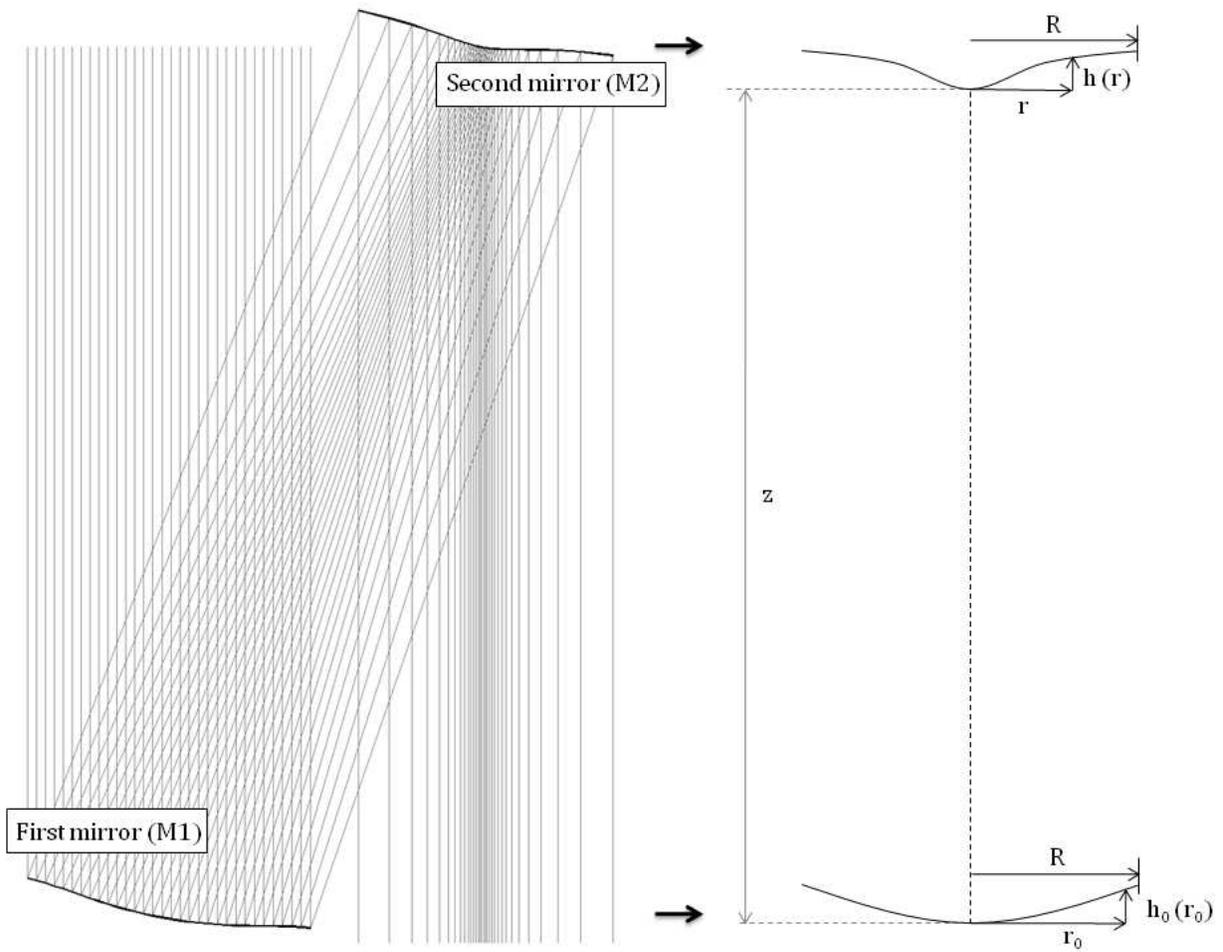}
\caption{The basic PIAA system, illustrated with rays: light entering the system is redirected by the first mirror to concentrate or spread out the light, and the second mirror corrects the path length so the wavefront has a uniform phase at the output as well as the input.  The mirror depths have been increased by a factor of 5 for clarity.  Coordinates related to arrangement of these mirrors are shown to the right.}
\label{fig:diag}
\end{center}
\end{figure}

A key problem in designing a PIAA system is determining what shapes the mirrors must take on to correctly redistribute the light.  Early PIAA work \mycitep{Tra03, Guy03} approached this from the perspective of ray optics, shaping the mirrors to redirect incoming rays into a desired amplitude distribution while maintaining constant pathlength for each ray.  However, in 2006, \mycitet{Van06} showed that this approach was insufficient; mapping of rays was insufficient to capture all of the physics seen in the full diffraction integral.  Rays \emph{do} allow for a straightforward visual interpretation of the remapping induced by PIAA, and the workings of a PIAA system using rays are shown in \myfig{fig:diag}.  (We note that, as an approximation, design is done in an on-axis configuration, despite the fact that the system geometry requires that mirrors must eventually be placed off-axis; the system layout is generally chosen to minimize the off-axis angle.)

In the following, we investigate an alternative approach to creating PIAA systems with pre- and post-apodizers.  Rather than ray optics, we start with the Huygens diffraction integral as set out in \mycitet{Van06}, with a uniform amplitude distribution.  We then consider the first term of an asymptotic expansion of the full Huygens diffraction integral---that is, apply the stationary phase approximation \mycitep{Bor99}---and show that the resulting approximation directly relates the output amplitude distribution and the mirror shapes, and additionally recovers the geometric optics ray-remapping relationship from the full diffraction integral for a uniform input amplitude.  In Sec. \ref{subsec:kikm}, we then use the stationary phase approximation to show how to remap non-uniform but radially-symmetric input amplitude distributions.

Our purpose in doing this is threefold: first, to show that a relation between input and output amplitude profiles and the mirror heights can be derived from the diffraction integral in a manner consistent with, but without recourse to, ray optics, and to be able to make the same sorts of statements about the amplitude distributions in a diffractive PIAA system that can be done with ray optics. We would note that the purpose here is \emph{not} to explicitly rederive the ray-optics mapping, though the equivalence does introduce a certain amount of convenience.

Second, the stationary-phase approximation gives a straightforward derivation for the relationships between the amplitude profiles at different planes that to our knowledge has not been shown in a ray-optics context for PIAA previously.

Third and finally, we wish to lay the groundwork for designs which explicitly minimize diffractive effects by appropriate choices of apodizers and mirror shapes.  The differences between the stationary phase approximation and the full diffraction integral can be expressed explicitly as higher-order terms in the same asymptotic expansion; the derivation of these terms falls beyond the scope of this current work, but their characteristics and general dependence on system parameters are known and discussed later in Sec. \ref{subsec:hot}, along with strategies for mitigating them.

Sec. \ref{subsec:kikf} and \ref{subsec:kmkf} show how the equations of Sec. \ref{subsec:kikm} may be inverted to allow the PIAA mirror shapes to be derived directly from specified input and output profiles, as well as how exactly the required profile on the input beam is set by the mirror shapes and the output profile.  In particular, the results of these sections allow one to choose apodizers and a desired output amplitude profile---based on whatever criteria which will optimize performance for the specific application---and determine exactly what shapes the mirrors must be to make this happen, a novel result for radially-symmetric PIAA systems.

\begin{figure}
\begin{center}
\includegraphics[width=6.5in]{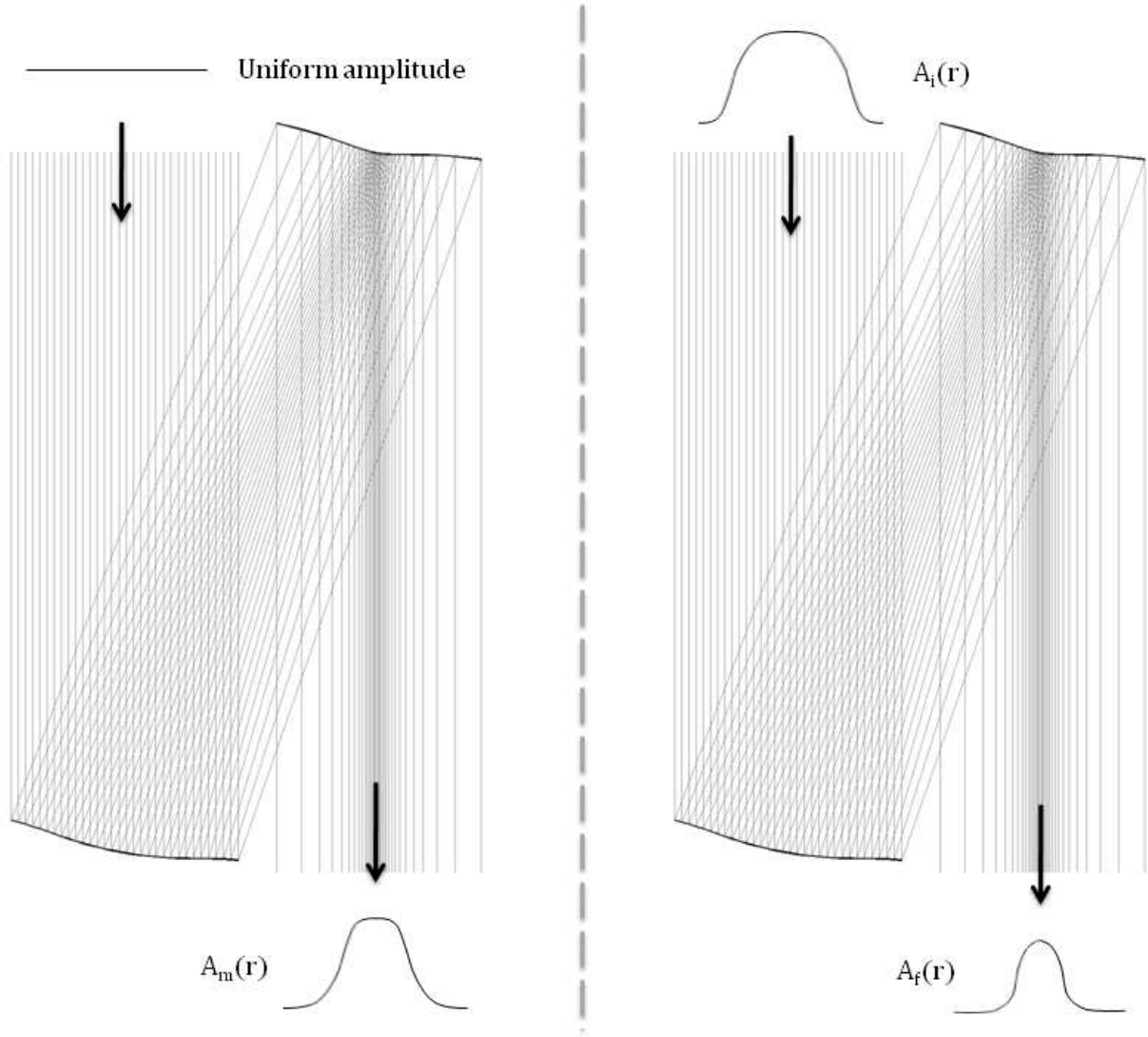}
\caption{A diagram showing the three amplitude profiles associated with a PIAA system. $A_{m}(r)$ is the output amplitude profile when the input amplitude profile to a set of mirrors is uniform, while for an arbitrary input profile $A_{i}(r)$ and the same set of mirrors, the output profile is $A_{f}(r)$.  As shown later, knowledge of any two of these profiles allows the third to be determined.}
\label{fig:amps}
\end{center}
\end{figure}

\subsection{The stationary phase approximation applied the diffraction integral}

For a radially-symmetric PIAA system, there are three different amplitude profiles to consider when propagating an on-axis wavefront:
\begin{itemize}
\item The amplitude profile of the wavefront incident on the first mirror, which we denote $A_{i}(r)$.
\item The amplitude profile directly following the second mirror, which we denote $A_{f}(r)$.
\item The `intrinsic' profile of the mirror pair; this profile would be produced by an incident plane wave when considering the remapping as ray-optics, without diffraction.  We denote this $A_{m}(r)$.
\end{itemize}
The amplitude profiles $A_{i}(r)$, $A_{f}(r)$, and $A_{m}(r)$ are shown in \myfig{fig:amps}.  These three profiles are not independent, but rather selecting any two determines the third; in the following we will show this and further show how to determine the third given the other two.  (A fourth profile from a post-apodizer, $A_{p}(r)$, may be included to reshape the amplitude profile further; this factor is solely multiplicative and does not factor into the remapping equations that follow.  See Sec. \ref{subsec:postap}.)

Assume we have a PIAA system with intrinsic profile $A_{m}(r)$, assumed real and positive.  Consider initially the output from a wavefront $E_{0}(x,y)$ incident on a PIAA system with no pre-apodizer; following the Huygens-Fresnel principle, the electric field at any point in the output plane can be considered to be a superposition of spherical waves from all points across the open aperture, which in Cartesian coordinates takes the form \mycitep{Van06}
\begin{equation} \label{Qint}
	E(x, y) = \int \int \frac{E_{0}(x,y)}{i \lambda Q(x_0, y_0, x, y)} e^{i \frac{2 \pi}{\lambda} Q(x_0, y_0, x, y)} \intd{x_0} \intd{y_0}.
\end{equation}
Here, $x_0$ and $y_0$ are coordinates in the plane of the input aperture, while $x$ and $y$ are coordinates in a plane just following the second mirror. In the case of an on-axis plane wave, $E_{0}(x,y) = 1$.  The optical path length $Q$ is defined as
\begin{align}
	Q(x_0, y_0, x, y) &= \sqrt{(x_0 - x)^2 + (y_0 - y)^2 + (z + h(r) - h_0(r_0))^2} + (h(r) - h_0(r_0)) \\
	&= \sqrt{r_0^2 + r^2 - 2 r_0 r \cos{(\theta -\theta_0)} + (z + h(r) - h_0(r_0))^2} + (h(r) - h_0(r_0)) \label{Q1},
\end{align}
where $(r_0, \theta_0)$ and $(r, \theta)$ are polar coordinates equivalent to $(x_0, y_0)$, and $(x, y)$, respectively. The shapes of the first and second mirror are given by $h_0(r_0)$ and $h(r)$, respectively, and defined by
\begin{equation} \label{h1}
	\dd{h_0(r_0)}{r_0} = \dd{h(r)}{r} = \frac{r_0(r) - r}{2 z},
\end{equation}
and
\begin{equation} \label{h2}
	h_0(0) = h(0) = 0.
\end{equation}
We note that this notation has been adjusted slightly from the definitions in \mycitet{Van06}, so that both $h_0$ and $h$ are between 0 and the maximum height of the mirror.  Here, unlike in \myeq{Q1}, $r_0$ in \myeq{h1} is not an independent variable from $r$; rather, we recall the equations for a single pair of radially-symmetric PIAA mirrors, which relate the coordinates $r_0$ on the first mirror to $r$ on the second (see \emph{e.g.} \mycitet{Van05}):
\begin{align}
	\dd{r_0}{r} &= \frac{r}{r_0}A_{m}(r)^2 \\
	\Rightarrow r_0(r) &= c\sqrt{2 \int_0^{r} s A_{m}(s)^2 ds}, \label{r0eq}
\end{align}
where $c$ is a normalization constant chosen so $r_0(R) = R$---that is, both mirrors have the same radii.

We also define the square root portion of $Q$ as $S(x_0, y_0, x, y)$, such that
\begin{align} \label{Sdef}
	S(x_0, y_0, x, y) &= \sqrt{(x_0 - x)^2 + (y_0 - y)^2 + (z + h(r) - h_0(r_0))^2}, \\
	Q(x_0, y_0, x, y) &= S(x_0, y_0, x, y) + (h(r) - h_0(r_0)).
\end{align}
($S$ will appear later in this derivation, and is an important factor in the S-Huygens approximation \mycitep{Van06, Bel06} which is used for diffraction calculations.)  The derivatives of $Q$ are equal to zero when $Q$, the path length, is a constant $Q_0$, which occurs (as shown in \mycitet{Van05}) when $r_0 = r_0(r)$, as defined by \myeq{r0eq}, and $\theta_0 = \theta$.

We approach the evaluation of \myeq{Qint} by using the stationary phase approximation (see appendix III of \mycitet{Bor99}), which allows integrals of a certain form to be approximated for large values of the parameter $k$:
\begin{equation} \label{gint}
	\int \int g(x_0, y_0) e^{i k f(x_0, y_0)} \intd{x_0} \intd{y_0} \approx \frac{1}{k}\frac{2 \pi i \sigma}{\sqrt{|\alpha \beta - \gamma^2|}} g(x_c, y_c) e^{i k f(x_c, y_c)},
\end{equation}
where $(x_c, y_c)$ are critical points where
\begin{equation}
  \dd{f}{x_0} = 0, \dd{f}{y_0} = 0,
\end{equation}
and
\begin{align} \label{abgs}
		\alpha &= \left.\frac{\partial^2 f}{\partial x_0^2} \right|_{x_c, y_c}, \\
		\beta &= \left.\frac{\partial^2 f}{\partial y_0^2} \right|_{x_c, y_c}, \\
		\gamma &= \left.\frac{\partial^2 f}{\partial x_0\partial y_0} \right|_{x_c, y_c},\\
		\sigma &= \left\{
		\begin{array}{cc}
		1 & \alpha \beta > \gamma^2, \alpha > 0, \\
		-1 & \alpha \beta > \gamma^2, \alpha < 0, \\
		i & \alpha \beta < \gamma^2.
		\end{array} \right.
\end{align}
We define $k$ ``large'' more explicitly as being such that $k f(x_0, y_0) \gg 2 \pi$, with the assumption that $g(x_0, y_0)$ is slowly varying and $f(x_0, y_0)$ is not constant over the region of integration.  Informally, the exponent must oscillate sufficiently quickly---and the amplitude $g(x_0, y_0)$ sufficiently slowly---that the contributions of neighboring regions tend to cancel each other out, except in the vicinity of certain critical points.

By applying \myeq{gint} to the equations governing the propagation through the PIAA system, we can determine the effect of incident wavefront profiles.  (This is, of course, an approximation; the stationary phase approximation is the first term in an asymptotic expansion of the integral in \myeq{Qint}.  The effects of higher-order terms in this expansion are discussed in Sec. \ref{subsec:hot}.)

We consider first the case of an on-axis plane-wave.  We can return to \myeq{gint} and rewrite its components as:
\begin{align}
k &= \frac{2 \pi}{\lambda}, \\
g(x_0, y_0) &= \frac{E_{0}(x,y)}{i \lambda Q(x_0, y_0, x, y)} = \frac{1}{i \lambda Q(x_0, y_0, x, y)} \,\mathrm{for}\,\mathrm{on}\mathrm{-}\mathrm{axis}\,\mathrm{plane}\,\mathrm{wave}, \label{gx0y0}\\
f(x_0, y_0) &= Q(x_0, y_0, x, y),
\end{align}
to make the left side identical to \myeq{Qint}; it remains only to evaluate the second derivatives of $Q$ to determine \myeq{abgs} and complete the approximation.  The algebra involved is extensive and is derived in full in \mycitet{Pue08} and \mycitet{Pue11a}, and we merely report the results: for $Q$, $S$, $h_0$, and $h$ as defined above,
\begin{align}
		\frac{\sigma}{\sqrt{|\alpha \beta - \gamma^2|}} &= \frac{1}{\sqrt{|\frac{1}{S(x_0(x_c), y_0(y_c), x_c, y_c)^2 A_{m}(r)^2}|}} \\
		&= A_{m}(r) S(x_0(x_c), y_0(y_c), x_c, y_c),
\end{align}
as both and $S$ and $A_{m}(r)$ are real and positive.

The right side of \myeq{gint} thus becomes:
\begin{align}
	\frac{1}{k}\frac{2 \pi i \sigma}{\sqrt{|\alpha \beta - \gamma^2|}} g(x_c, y_c) e^{i k f(x_c, y_c)} &= \frac{\lambda}{2 \pi} 2 \pi i A_{m}(r) \frac{S(x_0(x_c), y_0(y_c), x_c, y_c)}{i \lambda Q(x_0(x_c), y_0(y_c), x_c, y_c)} e^{2 \pi i Q(x_0(x_c), y_0(y_c), x_c, y_c)/\lambda }.
\end{align}
If we use the fact that $Q(x_0(x_c), y_0(y_c), x_c, y_c) = Q_0$, we reduce this to:
\begin{align}
    \frac{1}{k}\frac{2 \pi i \sigma}{\sqrt{|\alpha \beta - \gamma^2|}} g(x_c, y_c) e^{i k f(x_c, y_c)}	& = A_{m}(r) e^{2 \pi i Q_0/\lambda}\frac{S(x_0(x_c), y_0(y_c), x_c, y_c)}{Q_0}.
\end{align}
In general $h(r) - h_0(r_0) \ll S$, and so:
\begin{align}
\frac{S(x_0(x_c), y_0(y_c), x_c, y_c)}{Q_0} &\approx 1,  \\
\int \int \frac{1}{i \lambda Q(x_0, y_0, x, y)} e^{i \frac{2 \pi}{\lambda} Q(x_0, y_0, x, y)} \intd{x_0} \intd{y_0} &\approx A_{m}(r) e^{2 \pi i Q_0/\lambda}. \label{reQ}
\end{align}
and so the output wavefront for a plane wave is the desired intrinsic remapping $A_{m}(r)$ multiplied by a constant phase.  As described above, $A_{m}(r)$ is the output amplitude distribution for ray-optics propagation through a PIAA system, and so, for the case of a uniform input amplitude, this reproduces the results of geometric optics \mycitep{Van05}.

The substitution for $E_{0}(x,y)$ in \myeq{gx0y0} remains the only deviation from generality in the description of the input amplitude profile.  We will capitalize on this to relate the amplitude profiles at different planes without requiring any recourse to ray optics; this is the major strength of the stationary phase approximation in this application.

\subsection{Case 1: known $A_{i}(r)$, $A_{m}(r)$} \label{subsec:kikm}

Suppose the incident wavefront $E_{0}(x,y)$ in \myeq{Qint} is not a plane wave, but has instead has some radially-symmetric apodization profile $A_{i}(r)$ with $r = \sqrt{x^2 + y^2}$ and zero phase:
\begin{equation} \label{Qint2}
	E(x, y) = \int \int \frac{A_{i}(r)}{i \lambda Q(x_0, y_0, x, y)} e^{i \frac{2 \pi}{\lambda} Q(x_0, y_0, x, y)} \intd{x_0} \intd{y_0}.
\end{equation}
Then
\begin{align}
g(x, y) &= \frac{A_{i}(r)}{i \lambda Q}, \\
g(x_c, y_c) &= \frac{A_{i}(r_0(r))}{i \lambda Q_0},
\end{align}
and \myeq{gint} becomes
\begin{align}
	\frac{1}{k}\frac{2 \pi i \sigma}{\sqrt{|\alpha \beta - \gamma^2|}} g(x_c, y_c) e^{i k f(x_c, y_c)} &= \frac{\lambda}{2 \pi} 2 \pi i A_{m}(r) S(x_0(x_c), y_0(y_c), x_c, y_c) \frac{A_{i}(r_0(r))}{i \lambda Q_0} e^{2 \pi i Q_0/\lambda} \\
		& = A_{m}(r) A_{i}(r_0(r)) \frac{S(x_0(x_c), y_0(y_c), x_c, y_c) e^{2 \pi i Q_0/\lambda}}{Q_0}, \label{A1A2}
\end{align}
and thus, given $A_{m}(r)$ and $A_{i}(r)$, we can write
\begin{equation} \label{Afdef}
    A_{f}(r) = A_{m}(r) A_{i}\left(c\sqrt{2 \int_0^{r} s A_{m}(s)^2 ds}\right).
\end{equation}
Note the difference between this and the product of the two remapping functions $A_{i}(r)$ and $A_{m}(r)$; a pre-apodization does \emph{not} propagate straight through a PIAA coronagraph but is remapped to a certain extent.

\subsection{Case 2: known $A_{i}(r)$, $A_{f}(r)$} \label{subsec:kikf}

Next, we consider the case of a known $A_{i}(r)$ and $A_{f}(r)$.  This is a design problem for a mirror pair: given a pre-apodizer, designed to mitigate ringing, and a final apodization profile that will provide necessary contrast, what should the intrinsic profile $A_m(r)$ be?  To approach this, we start from \myeq{Afdef} and change variables:
\begin{align} \label{matchingCondition}
A_{f}(r) &= A_{m}(r) A_{i}\left(c\sqrt{2 \int_0^{r} s A_{m}(s)^2 ds}\right) = A_{m}(r)A_{i}(r_{0}(r)), \\
r_0(r) &= \sqrt{2 \int_{0}^{r} \sigma A_{m}(\sigma)^2 \intd{\sigma}},  \label{r0}\\
\dd{r_0}{r} &= \frac{r}{r_0}A_{m}(r)^2. \label{dr0}
\end{align}
As all apodizations will be $\geq 0$, we can square both sides and change variables to convert the entire system to functions of $r$ and $r_0$:
\begin{equation}
A_{f}(r)^2 = \dd{r_0}{r}\frac{r_0}{r}A_{i}(r_{0}(r))^2,
\end{equation}
and so $r_0(r)$ must satisfy the differential equation:
\begin{equation} \label{dr0dr}
\dd{r_0}{r} = \frac{r A_{f}(r)^2}{r_0 A_{i}(r_{0}(r))^2}.
\end{equation}
This is a separable first-order ordinary differential equation, whose general solution is well-known (see \emph{e.g.} \mycitet{Arf84}):
\begin{equation} \label{solution}
\int_0^r \alpha A_{f}(\alpha)^2 \intd{\alpha} = \int_0^{r_0} \beta A_{i}(\beta)^2 \intd{\beta}.
\end{equation}
These cumulative functions on $A_{i}$ and $A_{f}$ may be computed (numerically, in practice), and the $r_0(r)$ associated with any $r$ determined from the \myeq{solution} equality.  As the apodization functions are both real and nonnegative, the cumulative integrals will be nondecreasing, which simplifies the numerics; we also recall that for a pair of mirrors with identical radius $R$ which preserve throughput, $r = r_0(r)$ at $r = 0$ and $r = R$, which should be useful as a check.  $A_{m}(r)$ may subsequently be derived using \myeq{matchingCondition}:
\begin{equation}
A_{m}(r) = \frac{A_{f}(r)}{A_{i}(r_{0}(r))}.  \label{Amdef}
\end{equation}

We note that knowledge of $A_m(r)$ directly determines the heights of the PIAA mirrors via \myeq{h1} and \myeq{h2}, and thus we can use \myeq{solution} and \myeq{Amdef} to analytically determine the mirror shapes required to take any amplitude profile to any other---a novel result in PIAA design.  This approach permits a number of useful applications, such as compensating for pre-apodizer effects and chaining together pairs of PIAA mirrors; we return to these possibilities in Sec. \ref{sec:app}.

\subsection{Case 3: known $A_{m}(r)$, $A_{f}(r)$} \label{subsec:kmkf}

Finally, we consider the case of a known $A_{m}(r)$ and $A_{f}(r)$.  This is a design problem for a pre-apodizer: given a mirror pair, what must the input apodization be to get a desired output?

To approach this, we return to \myeq{r0eq}:
\begin{align}
	r_0(r) &= c\sqrt{2 \int_0^{r} s A_{m}(s)^2 ds}, \label{r0invfunc}
\end{align}
and invert this equation (generally numerically, in practice) to find $r(r_0) \equiv r_0^{-1}(r_0)$.  As again $A_{m}(r)$ is real and nonnegative, the cumulative integral (and thus $r_0(r)$) will be nondecreasing  which will again simplify numerics.  (As a check, the resulting inverse function should satisfy $r_0^{-1}(r_0(r)) = r$ for $ 0 \leq r \leq R$.)

A change of variable in \myeq{Amdef} then gives that
\begin{equation}
A_{m}(r_0^{-1}(r)) = \frac{A_{f}(r_0^{-1}(r))}{A_{i}(r)},
\end{equation}
and thus
\begin{equation} \label{Aidef}
A_{i}(r) = \frac{A_{f}(r_0^{-1}(r))}{A_{m}(r_0^{-1}(r))}.
\end{equation}

\myeq{r0invfunc}, and \myeq{Aidef}, will be valid as long as $A_{m}(r) > 0$; we can easily show the mapping from $r$ to $r_0$ must be invertible under this assumption by assuming the converse.  Suppose there exist an $r_1$ and $r_2$ such that $0 \leq r_1 < r_2$ and $r_0(r_1) = r_0(r_2)$.  This implies that
\begin{align}
    c\sqrt{2 \int_0^{r_1} s A_{m}(s)^2 ds} &= c\sqrt{2 \int_0^{r_2} s A_{m}(s)^2 ds} \\
    \Rightarrow \sqrt{\int_0^{r_1} s A_{m}(s)^2 ds} &= \sqrt{\int_0^{r_2} s A_{m}(s)^2 ds}.
\end{align}
Since $A_{m}(r)$ is real, and all $r$ are nonnegative, $s A_{m}(s)^2 \geq 0$ and so
\begin{align}
    \int_0^{r_1} s A_{m}(s)^2 ds &= \int_0^{r_2} s A_{m}(s)^2 ds. \label{contraEq}
\end{align}
Since $r_1 < r_2$,
\begin{align}
    \int_0^{r_2} s A_{m}(s)^2 ds &= \int_0^{r_1} s A_{m}(s)^2 ds + \int_{r_1}^{r_2} s A_{m}(s)^2 ds \\
    \Rightarrow \int_0^{r_2} s A_{m}(s)^2 ds - \int_0^{r_1} s A_{m}(s)^2 ds = 0 &= \int_{r_1}^{r_2} s A_{m}(s)^2 ds,
\end{align}
by \myeq{contraEq}.  However, if $A_{m}(r) > 0$,
\begin{align}
\int_{r_1}^{r_2} s A_{m}(s)^2 ds &\geq \left(\min{A_{m}(r)}\right)^2 \int_{r_1}^{r_2} s ds \\
\int_{r_1}^{r_2} s ds &= \frac{r_2^2}{2} - \frac{r_1^2}{2} > 0 \\
\Rightarrow \int_{r_1}^{r_2} s A_{m}(s)^2 ds &\geq \left(\min{A_{m}(r)}\right)^2 \left(\frac{r_2^2}{2} - \frac{r_1^2}{2}\right) > 0,
\end{align}
and we have a contradiction, thus the mapping will be invertible under these conditions.

\subsection{Higher-order terms in the stationary phase approximation} \label{subsec:hot}

\begin{figure}
\begin{center}
\includegraphics[width=5.25in]{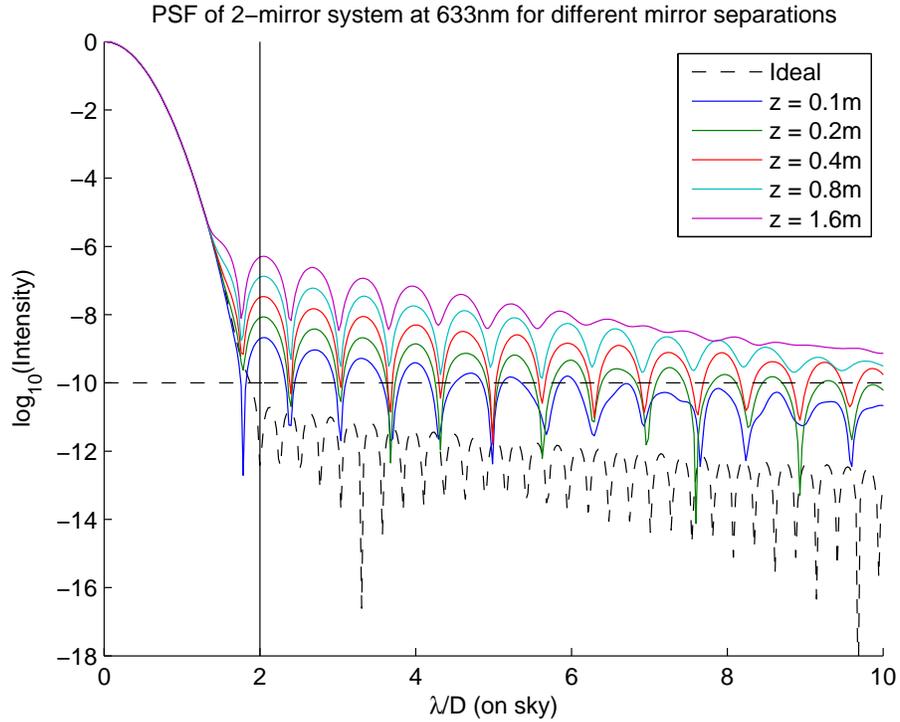}
\caption{The effect of varying mirror distance on the image plane performance of a PIAA system with pre- and post-apodizers.  Each solid curve shows the PSF of a 2-mirror system with a different spacing between the mirrors, computed with the S-Huygens approximation \mycitep{Bel06} to the full diffraction integral.  The dashed curve is the PSF produced by the ideal $A_{f}$ of the system multiplied by a post-apodizer $A_{p}$.}
\label{fig:beta}
\end{center}
\end{figure}

\begin{figure}
\begin{center}
\includegraphics[width=5.25in]{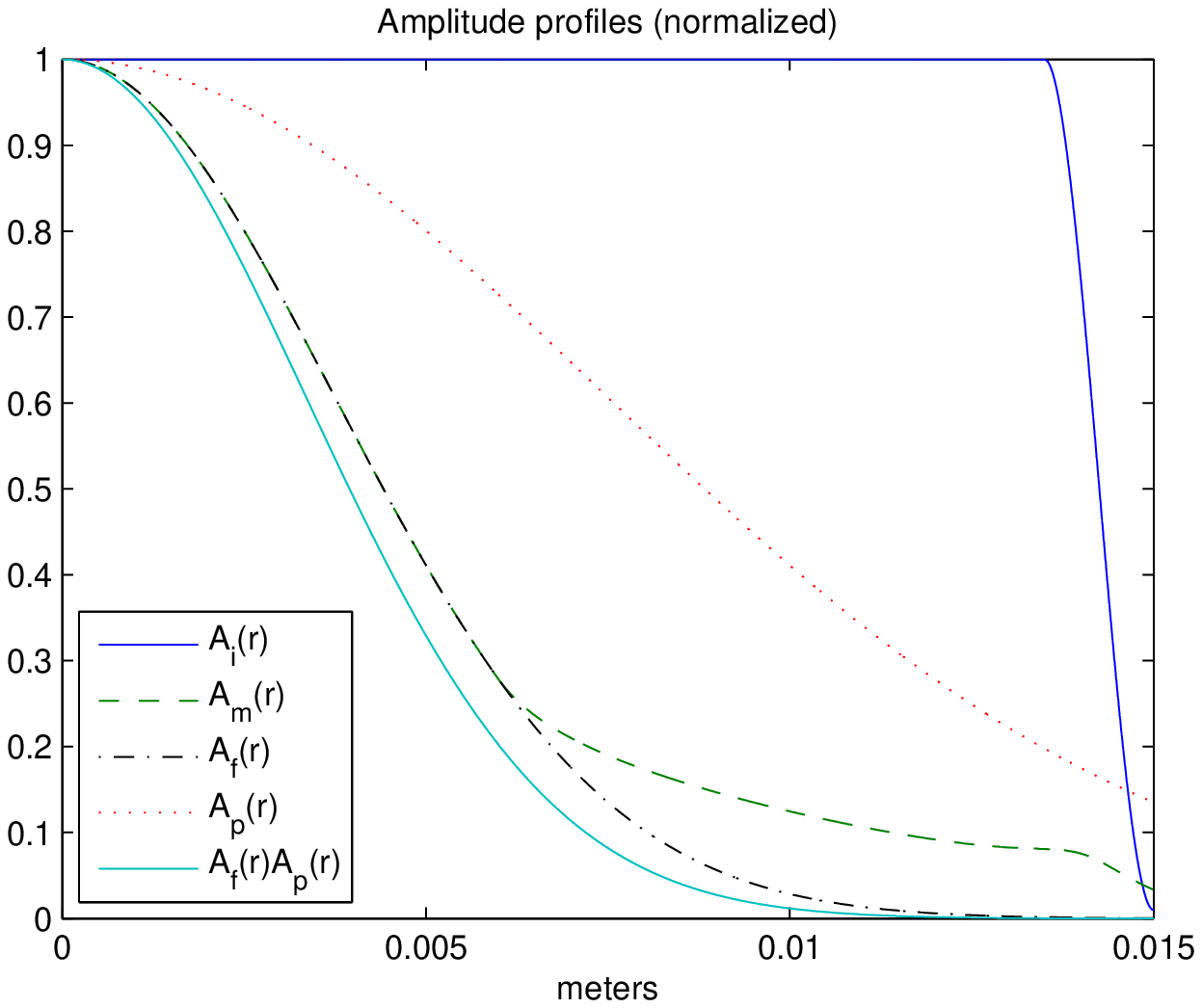}\\
\includegraphics[width=5.25in]{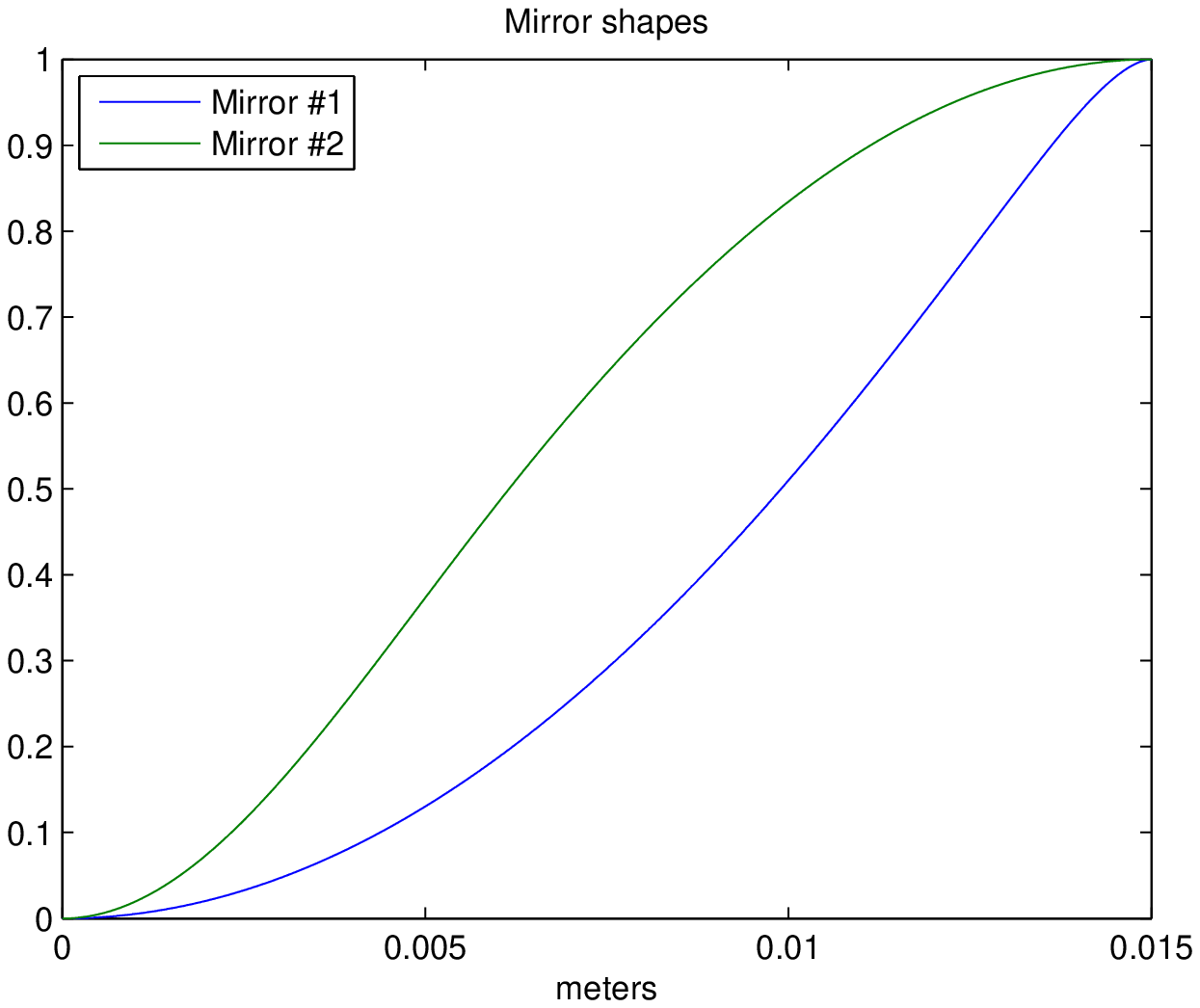}
\caption{\emph{Top.}  The amplitude profiles of the 2-mirror system in \myfig{fig:beta}, along with the apodization profile of the post-apodizer and the final profile produced by the system.  All are normalized to 1 at the center.  \emph{Bottom.} The shape of the two mirrors, normalized to unit height on the outer edge.  All mirror designs use an identical $A_m(r)$ and mirror radius, which gives identical mirror shapes scaled to greater or lesser sag following \myeq{h1} and \myeq{h2}.}
\label{fig:beta2}
\end{center}
\end{figure}

\myeq{Afdef}, \myeq{Amdef}, and \myeq{Aidef} are independent of any wavelengths, distances, and aperture sizes, and yet full diffraction modeling of a PIAA system shows the system can perform quite differently as these parameters are varied.  \myfig{fig:beta} shows an examples of this, using mirror spacing as a variable.  This system used both pre- and post-apodizers, which were held constant as $z$ was changed, along with pair of mirrors which correspond to the same $A_{m}(r)$ for all values of $z$, and these can be seen in \myfig{fig:beta2}. (This means the mirrors will always have the same shape, but scaled up and down to take the maximum height of the mirror from a low of $21\mu$m at $z = 1.6$m to a high of $350\mu$m at $z = 0.1$m.) The mirror profiles were determined with \myeq{solution} and \myeq{Amdef}, using the polynomial pre-apodizer outlined later in Sec. \ref{subsec:md}, and both a Gaussian $A_{f}(r)$ and post-apodizer $A_{p}(r)$.

It is well-established \mycitep{Van06} that the contrast from a pair of mirrors alone will be overwhelmed by diffraction from the first PIAA mirror.  The inclusion of a pre-apodizer and post-apodizer to a mirror pair is necessary but still \emph{not} sufficient to produce a system with good performance; diffraction modeling must be run on the actual system configuration to verify the system performs adequately.

The explanation for why the system can perform so differently is that there always remain additional wavelength- and separation-dependent diffraction effects \mycitep{Van06} which have not been accounted for in the design process.  These residual diffraction effects correspond to higher-order terms in the asymptotic expansion (of which the stationary phase approximation is the leading order) and must be accounted for---and compensated with apodizers and system properties---to properly determine the expected performance of a PIAA system.  A full discussion of these terms and their effect on PIAA systems is beyond the scope of this paper, but we can draw on related analysis from the laser beam shaping community to present a few salient considerations.

The key parameter for quality of a beam-shaping optical system is $\beta$ \mycitep{Dic00}, which in PIAA terms can be defined as
\begin{equation}
\beta = \frac{2 \pi R^2}{\lambda z}.
\end{equation}
The larger this parameter becomes, the closer a beam-shaping system (or a PIAA system) comes to the first term in the stationary phase approximation.  For small $\beta$ performance will be hindered; for sufficiently small $\beta$, it becomes mathematically impossible to get a beam shaping system that works even approximately, as this would violate the uncertainty principle for Fourier transforms \mycitep{Rom96}, although this will generally not be an issue for most PIAA systems. In practice, this means that building larger mirrors or lenses and placing them closer together will reduce diffraction effects; it also suggests that if $\beta$ is sufficiently large at the longest wavelength of operation, the PIAA system performance will be wavelength-independent.  (At the level of design, at least; a real system will inevitably have wavelength-dependent performance due to aberrations in the optics, but it should not be limited by diffraction.)

The falloff of higher-order terms depends on the continuity class of $Q(x_0, y_0, x, y)$.  For an infinitely differentiable $Q$, the higher order terms will fall off as $1/\beta^2$; however, if $Q$ has a discontinuity in its third derivative, or $A_{i}(r)$ has a discontinuity in its first derivative, then the rate of convergence will drop to $1/\sqrt{\beta}$  \mycitep{Dic00}. Discontinuities in lower-order derivatives will reduce the convergence further.  The input field is a plane wave with a discontinuity at the edge of the aperture, which immediately suggests that inserting an apodizer (see Sec. \ref{subsec:preap}) to produce a continuous $A_{i}(r)$ will improve convergence.  (This also suggests that the choice of $A_m(r)$ may affect convergence properties and thus exacerbate residual diffraction---something to keep in mind for design.)

Compensating for these diffraction effects \emph{a priori} is nontrivial, though a hope for future analysis is that the dependence of those terms on $A_{i}(r)$ and $A_{m}(r)$ and their derivatives can be characterized and used to set conditions on these functions which can minimize or null the largest terms in the asymptotic expansion.  (Ideally, this would allow PIAA systems to be designed which naturally minimize diffractive effects.)  One fairly simple solution to mitigate some of these effects is to note that while the first mirror's main purpose is to redirect the light, the purpose of the second is to correct the phase of the wavefront so that the output has a uniform phase.  The diffraction effects above will introduce phase errors which reduce the contrast in the final image plane; to compensate, we can model the residual phase at a wavelength near the center of our band of operation and add half this value to the surface of the second mirror.  Some smoothing may be applied if desired, as the residual phase may have high-frequency components.  This approach is also taken in \mycitet{Guy10}, to correct similar phase effects.  This simple correction can reduce the residual light near the inner working angle by two or more orders of magnitude; the examples in the next section use this as a final adjustment where noted.

\section{Applications} \label{sec:app}

\subsection{Post-apodizer design} \label{subsec:postap}

The work of \mycitet{Van06} shows that a PIAA system alone will be limited to $\approx10^{-5}$ contrast by diffraction considerations; one method investigated in that paper and in \mycitet{Plu06} to press beyond this limit was the use of apodizers preceding (a \emph{pre-apodizer}) and following (a \emph{post-apodizer}) the mirrors.  The post-apodizer would do some of the beam shaping, but at the expense of throughput.  (Pre-apodizers will be discussed in the next section.)

The choice of the post-apodizer is a free variable; switching out post-apodizers is equivalent to changing the final profile of the system. The profile is selected to produce sufficient contrast at the image plane in conjunction with the PIAA mirrors.  \myeq{Afdef} can be modified slightly to split $A_f(r)$ into a post-apodizer component $A_{p}(r)$ and the final output of the combined system $A_{\mathrm{tot}}(r)$:
\begin{align}
A_{\mathrm{tot}}(r) &= A_{p}(r) A_{m}(r) A_{i}\left(c\sqrt{2 \int_0^{r} s A_{m}(s)^2 ds}\right). \label{withPost}
\end{align}

\subsection{Pre-apodizer design} \label{subsec:preap}

One difficulty with PIAA systems is the that the beam is expected to fill the mirror; this produces diffraction from the edges, which (as discussed above) will limit the performance of the system.  To compensate, we design preapodizers, which alter the discontinous step at the edge of the mirror into a smoother function.  In general, it is better to design the pre-apodization in advance, since diffraction mitigation introduces an additional constraint on $A_{i}(r)$ not found in $A_{m}(r)$ or $A_{f}(r)$; nonetheless, for completeness, we provide an example of how to retrofit a pair of mirrors with a pre-apodizer.

Suppose we wish to make a PIAA system which is designed to produce, as an output, a Gaussian apodization. (\mycitet{Bel06a} shows these can be used to make high-contrast, small IWA PIAA systems.)
\begin{equation} \label{pa1}
 	A_{f}(r) = c_1 e^{-a \left(\frac{r}{R}\right)^2}, a > 0.
\end{equation}
Additionally, let us suppose we wish to do this with a mirror set whose intrinsic profile is a Gaussian which falls off more slowly:
\begin{equation}
 	A_{m}(r) = c_2 e^{-b \left(\frac{r}{R}\right)^2}, 0 < b < a.
\end{equation}
The choice of a Gaussian, while providing adequate performance, is neither unique nor particularly special; we choose this particular profile for pedagogical reasons, as it produces an analytic representation for the preapodizer.  Using \myeq{r0eq}, we find $r_0(r)$,
\begin{align}
	r_0(r) &= \sqrt{2 \int_0^{r} \sigma c_2^2 e^{-2 b \left(\frac{r}{R}\right)^2}} \\
	&= c_2 \sqrt{\frac{R^2}{2 b}\left[1 - e^{-2 b \left(\frac{r}{R}\right)^2}\right]},
\end{align}
and we can write the inverse mapping for this case, as well:
\begin{align} \label{r0inv}
	r_0^{-1} = r(r_0) = \sqrt{\frac{-R^2}{2 b}\log{\left( 1 - \frac{2 b r_0^2}{R^2 c_2^2}\right)}}.
\end{align}
We can thus use \myeq{Aidef} to write the required input apodization:
\begin{align}
A_{i}(r) &= \frac{A_{f}(r_0^{-1}(r))}{A_{m}(r_0^{-1}(r))} \\
&= \frac{c_1}{c_2} e^{-(a-b) \left(\frac{r_0^{-1}(r)}{R}\right)^2} \\
&= \frac{c_1}{c_2} e^{-(a-b) \left(\frac{\sqrt{\frac{-R^2}{2 b}\log{\left( 1 - \frac{2 b r^2}{R^2 c_2^2}\right)}}}{R}\right)^2} \\
&= \frac{c_1}{c_2} e^{\frac{a-b}{2 b}\log{\left( 1 - \frac{2 b r^2}{R^2 c_2^2}\right)}}\\
&= \frac{c_1}{c_2} e^{\log{\left( 1 - \frac{2 b r^2}{R^2 c_2^2}\right)^{\frac{a-b}{2 b}}}}\\
\Rightarrow A_{i}(r) &= \frac{c_1}{c_2} \left( 1 - \frac{2 b r^2}{R^2 c_2^2}\right)^{\frac{a-b}{2 b}}. \label{inputAp}
\end{align}

We note, however, that most inverse mappings are not so fortunate as to have closed-form solutions as in \myeq{r0inv}; the inverse mapping must generally be done numerically.  For example, if we even add a small pedestal to the intrinsic profile to improve the manufacturability of the mirror, as suggested by \mycitet{Plu06}:
\begin{equation}
	A_{m,p}(r) = c_2 \frac{e^{- b \left(\frac{r}{R}\right)^2} + \beta}{1 + \beta},
\end{equation}
the mapping function becomes
\begin{equation}
	r_{0,p}(r) = c_2^2 \sqrt{\frac{\beta^2}{(1+\beta)^2} r^2 + \frac{4 \beta R^2}{(1+\beta)^2 b}
	(1- e^{- b \left(\frac{r}{R}\right)^2}) + \frac{R^2}{(1+\beta)^2 b}(1- e^{- 2 b \left(\frac{r}{R}\right)^2})},
\end{equation}
and the inverse mapping function is no longer expressible in closed form.

\subsection{Mirror design} \label{subsec:md}

One difficulty with pre-apodizer design is that a pre-apodizer has to meet an additional requirement that is not necessarily imposed on the other system optics: it needs to be entirely open across the majority of its aperture to keep throughput high, and yet has to approach zero transmission near the edge to prevent ringing from the discontinuity in field imposed by the mirror edge.  The pre-apodizer profile should not go all the way to zero, though, as this drives $A_{m}(r)$ to infinity in \myeq{Amdef}; in addition, making the profile too close to 0 or 1 at any point creates difficulties in manufacturing the apodizer.  (If the profile is intended to be truncated at some point, it is better to include this truncation in the profile $A_{i}(r)$ directly so that its effects can be compensated for.)

For these reason, it can be more advantageous to choose a desired output profile and a pre-apodizer which meets these requirements and design a mirror which connects the two.  Consider, for example, a simple pre-apodizer with an open center and a polynomial falloff near the edge:
\begin{align}
    A_{i}(r) &= \left\{
    \begin{array}{lc}
    1, & r \leq R_p \\
    a + 3 (b-a) \left(\frac{r - R_p}{R-R_p}\right)^2 + 2 (a-b) \left(\frac{r - R_p}{R-R_p}\right)^3, & R_p \leq r \leq R \\
    0, & r \geq R.
    \end{array} \right.
\end{align}
The polynomial in the region $R_p \leq r \leq R$ is chosen so that $A_{i}(R_p) = a$, $A_{i}(R) = b$, and $\dd{A_{i}(r)}{r}|_{r = R_p^{+}} = \dd{A_{i}(r)}{r}|_{r = R^{-}} = 0$.  Generally, we would prefer $a$ to be as close to $1$ and $b$ as close to $0$ as manufacturing permit.  This is not to imply that this is necessarily the best choice of pre-apodizer, but it has one significant computational advantage for design: the integral in \myeq{solution} can be done analytically, which allows the right side of \myeq{solution} to be implemented directly as a function instead of repeatedly determined by numerical integration.  (The solution, which is a large and not particularly instructive 8th-order polynomial, is not shown here.)

\begin{figure}
\begin{center}
\includegraphics[width=5.25in]{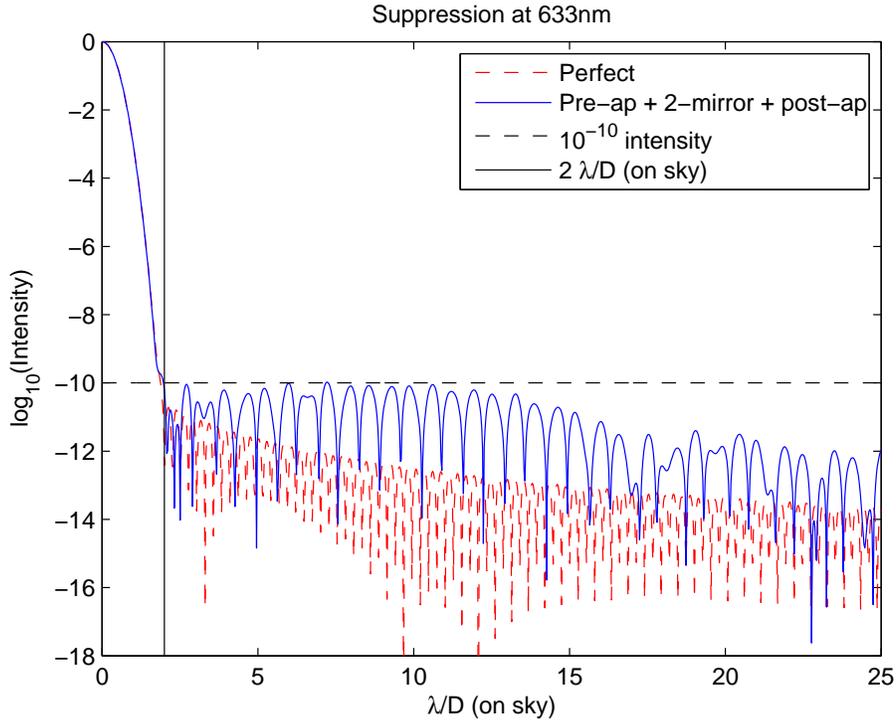}
\caption{A comparison between the PSFs from a set of amplitude profiles derived from the equations in Section \ref{sec:3f} (``Perfect'') and from the full diffractive treatment on the corresponding mirrors.  The two-mirror system has a preselected pre- and post-apodizer, and the mirrors designed with \myeq{Amdef}.  The second mirror is adjusted to compensate for phase errors, following \S~\ref{subsec:hot}.}
\label{fig:md1}
\end{center}
\end{figure}

If we then select a final profile and a post-apodizer, \myeq{Amdef} gives our mirror pair immediately.  An example of this is shown in \myfig{fig:md1}, using a Gaussian final profile and post-apodizer; this system would achieve $10^{-10}$ contrast at $2 \lambda/D$ on sky.  The red curve is the idealized output expected from geometric optics, while the blue curve includes the effects of diffraction.  The parameters in this example are given in \mytable{table:md1}; the post-apodizer was proportional to $e^{-2 (r/R)^2}$, and the final profile was proportional to $e^{-10 (r/R)^2}$.   Propagation through the system is performed using the S-Huygens approximation to the Rayleigh-Sommerfeld integrals following the prescription in \mycitet{Bel06}.  As discussed in Sec. \ref{subsec:hot}, the shape of the second mirror is adjusted to compensate for phase errors introduced by the higher-order terms neglected in taking the stationary-phase approximation.

\begin{table}
\begin{center}
\begin{tabular}{c|c}
\hline
Parameter & Value \\
\hline
$$a & 1 \\
$b$ & 0.01 \\
$R$ & 15mm\\
$R_{p}$ & 13.5mm\\
$z$ & 250mm\\
$\lambda$ & 633nm\\
\end{tabular}
\caption{Parameters used in \myfig{fig:md1}.} \label{table:md1}
\end{center}
\end{table}

\subsection{Example: Gaussian input beams}

During an astronomical observation of a distant star, the incident field will be negligibly close to a uniform-amplitude plane wave.  In laboratory PIAA experiments \mycitep{Bel09, Ker09, Guy10, Bal10}, however, the starlight is simulated with a laser, either directly or as the output of a single-mode fiber.  These beams, for lasers with high mode purity, will have Gaussian amplitude profiles, which effectively change the pre-apodization of the beam.   If that beam is not sufficiently oversized with respect to the PIAA mirrors, then the additional preapodization from the Gaussian beam will be remapped inward and affect the inner working angle of the coronagraph, at least in the laboratory setting.

Consider the case of a collimated laser beam with a single Gaussian mode, with a $\frac{1}{e}$-point in its amplitude at a radius $r_g$:
\begin{equation} \label{gauss}
    A_{\ell}(r) = c_{\ell} e^{-\left(\frac{r}{r_g}\right)^2},
\end{equation}
entering a PIAA system with a pre-apodizer, two mirrors, and a post-apodizer.  The Gaussian amplitude profile and the pre-apodizer combine to create an effective pre-apodization $A_{i}(r)A_{\ell}(r)$, which is remapped to $A_{i}(r_0(r))A_{\ell}(r_0(r))$ by the PIAA system.  Following \myeq{withPost}, the amplitude profile at the output becomes
\begin{align}
    A_{\mathrm{tot}}(r) &= A_{p}(r) A_{m}(r) A_{i}\left(r_0(r)\right) A_{\ell}(r_0(r)) \\
    &= A_{p}(r) A_{m}(r) A_{i}\left(c\sqrt{2 \int_0^{r} s A_{m}(s)^2 ds}\right) c_{\ell} e^{-\frac{2 c^2}{r_g^2} \int_0^{r} s A_{m}(s)^2 ds}.
\end{align}

\begin{figure}
\begin{center}
\subfigure{\includegraphics[width=3.25in]{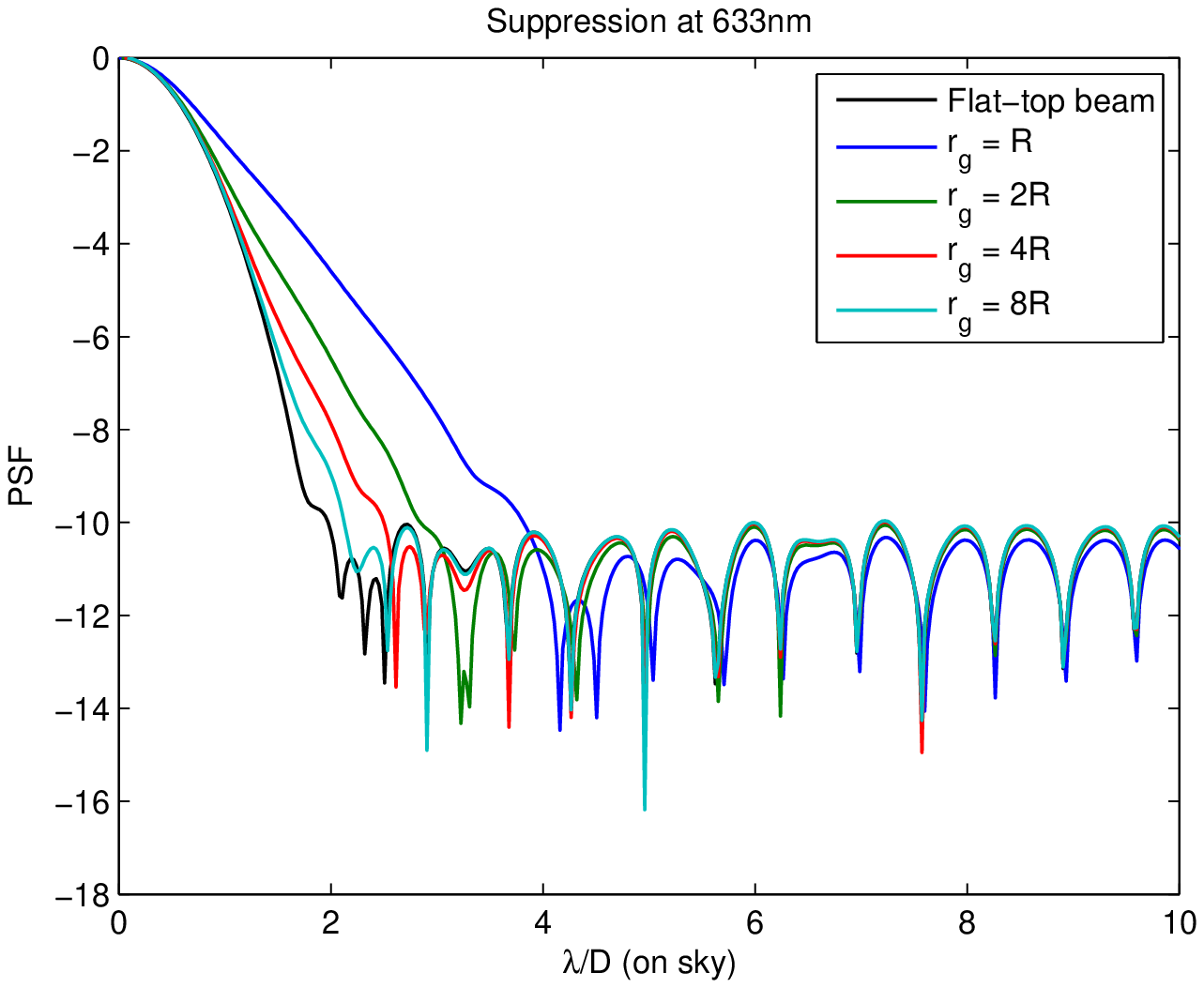}
\includegraphics[width=3.25in]{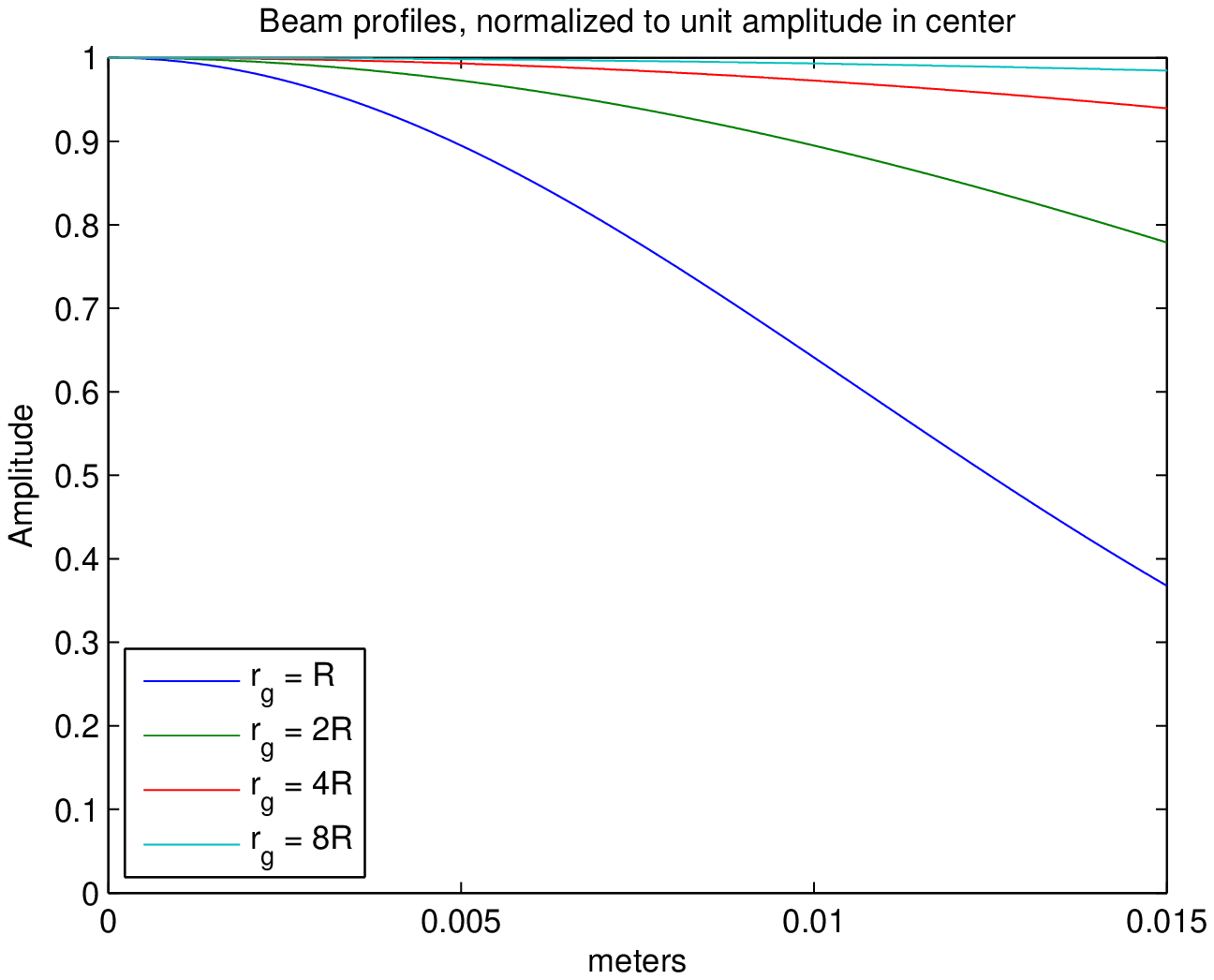}}
\caption{The effect of an incident beam with a Gaussian amplitude profile on the system shown in \myfig{fig:md1}. \emph{Left.} PSFs from the propagation of beams with Gaussian amplitude profiles $e^{-(r/r_g)^2}$ through a PIAA system. Here $R$ is the radius of the mirror. \emph{Right.} The associated beam amplitudes, normalized to unit amplitude in the center.}
\label{fig:rg}
\end{center}
\end{figure}

The remapped profile $A_{\ell}(r_0(r))$ may have non-trivial effects on the inner working angle; \myfig{fig:rg} shows the two-mirror system from Sec. \ref{subsec:md} and \myfig{fig:md1} with 4 inputs of varying beamwidth.  The $r_g$ are shown in units of the mirror radius $R$, and propagation is again done using the S-Huygens approximation.  (We note that if the remapping of the input is incorrectly ignored, the Gaussian profile $A_{\ell}(r)$ will multiply the $A_{f}(r)$ directly; in the case of \myfig{fig:md1}, this would produce yet another Gaussian profile with an even faster falloff.)

If the beam is not significantly oversized with respect to the PIAA mirrors, the effective change in the pre-apodization introduces significant amounts of light near the inner working angle, pushing it outward.  This effect should be taken into consideration when preparing an input beam or modeling its performance in laboratory settings.

\section{Multi-PIAA systems} \label{sec:4m}

One of the major problems with the use of a postapodizer is the ensuing drop in throughput, which removes one of the main advantages of PIAA.  In addition, they sacrifice some inner working angle, as a post-apodizer does not get the magnification bonus of a pair of PIAA mirrors.  To compensate for this, we can consider using a second set of PIAA mirrors to do a remapping, with the intrinsic profile of the second set equivalent to the apodization profile of the postapodizer.

To create a system with two sets of PIAA mirrors and no postapodizer, we first choose an overall final profile $A_{f,sys}(r)$ for the system, and an intrinsic $A_{m,2}(r)$ for the second set.  We solve \myeq{Aidef} to produce $A_{i, 2}(r)$ for the second set, and as the field out of the first pair of mirrors must match the field into the second, we have the matching condition:
\begin{equation} \label{matching}
    A_{i,2}(r) = A_{f, 1}(r).
\end{equation}
We choose a preapodizer $A_{i,1}(r)$ to minimize ringing in the system from the edges of the input aperture, and finally use \myeq{Amdef} to find $A_{m,1}(r)$.  The second mirror in the first set (and possibly the second set, as well) may be adjusted for phase errors.  As these equations are derived from the leading term of the expansion of the full diffraction integral, and the appropriate choice of pre-apodizer and mirror size/spacing will minimize effects of higher-order terms, we can chain four mirrors together in a way that can provide adequate contrast even with all diffraction effects included.

\begin{figure}
\begin{center}
\includegraphics[width=4.75in]{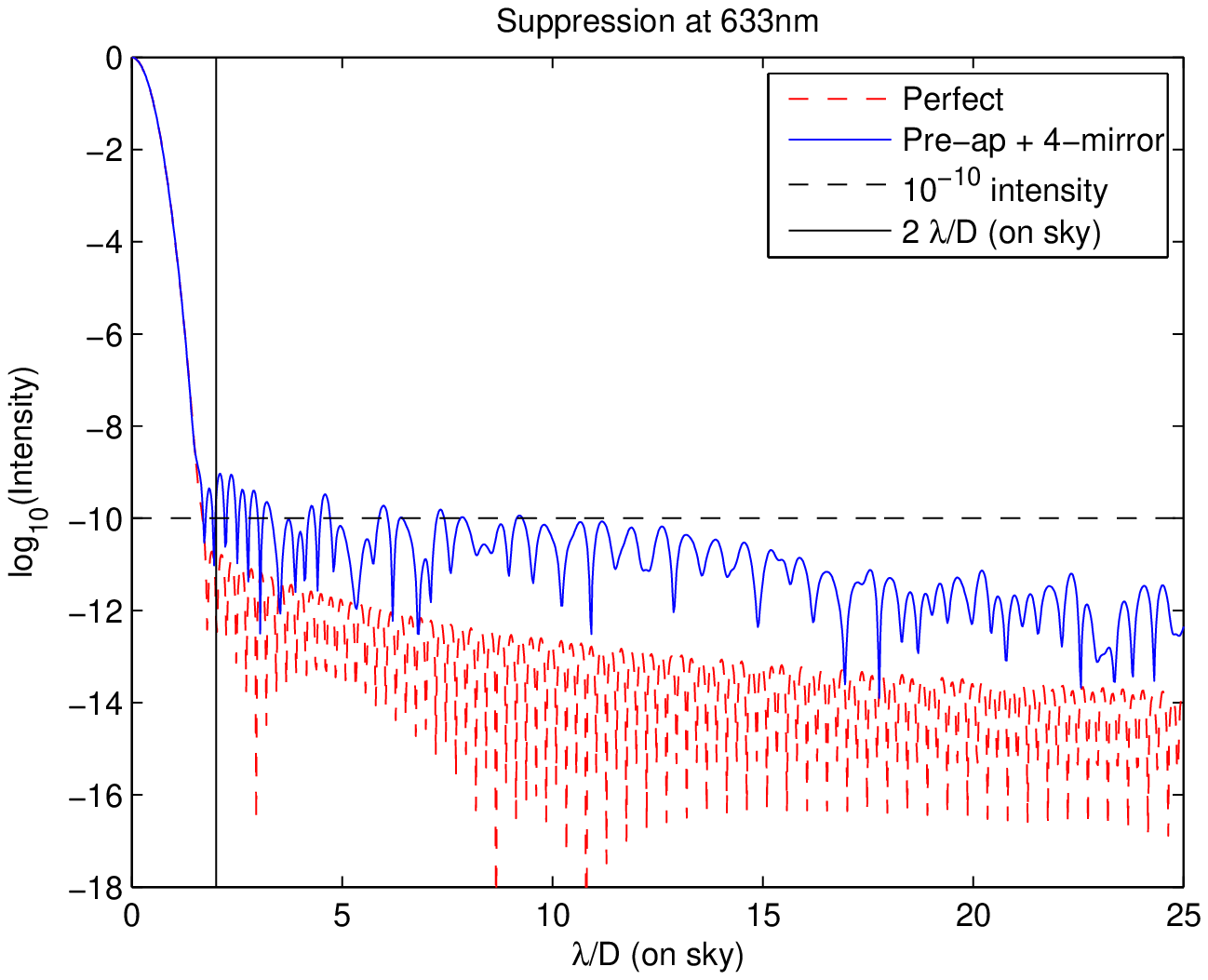}\\
\includegraphics[width=4.75in]{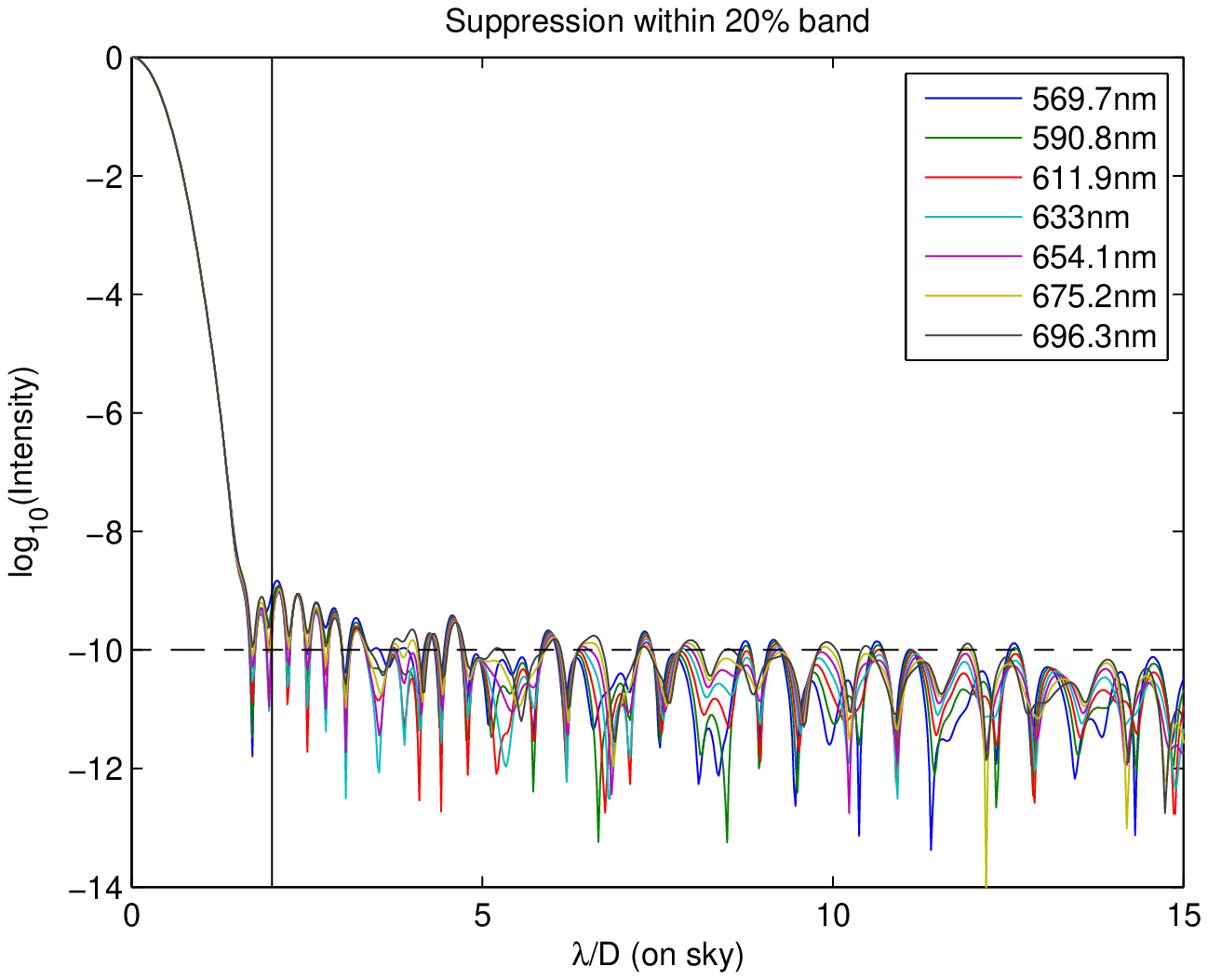}
\caption{\emph{Top.} A comparison of PSFs between a four-mirror system with a second set of mirrors chosen to mimic the post-apodizer used for \myfig{fig:md1}, and the corresponding idealized amplitude profiles (``Perfect''). \myeq{matching} is used to ensure the two pairs of mirrors couple together.  The second mirror in the first set is again adjusted to compensate for phase errors, following Sec. \ref{subsec:hot}.  \emph{Bottom.} The PSFs produced by the four-mirror system across a $20\%$ bandpass; the variation in the PSF is quite small.}
\label{fig:mm1}
\end{center}
\end{figure}

As an example, we can recreate the system given in \ref{subsec:md} using four mirrors instead of using a two mirrors and a post-apodizer.   The parameters chosen are identical to the ones in \mytable{table:md1}, and the second mirror has $A_{m,2} \propto e^{-2 (r/R)^2)}$ to match the post-apodizer.  The final system profile is still proportional to $e^{-10 (r/R)^2}$.  The resulting intensities are shown at top in \myfig{fig:mm1}; as with \myfig{fig:md1}, the red curve is from geometric optics, while the blue curve includes diffraction using S-Huygens. Broadband performance is shown in \myfig{fig:mm1} at bottom.

\begin{figure}
\begin{center}
\includegraphics[width=6.5in]{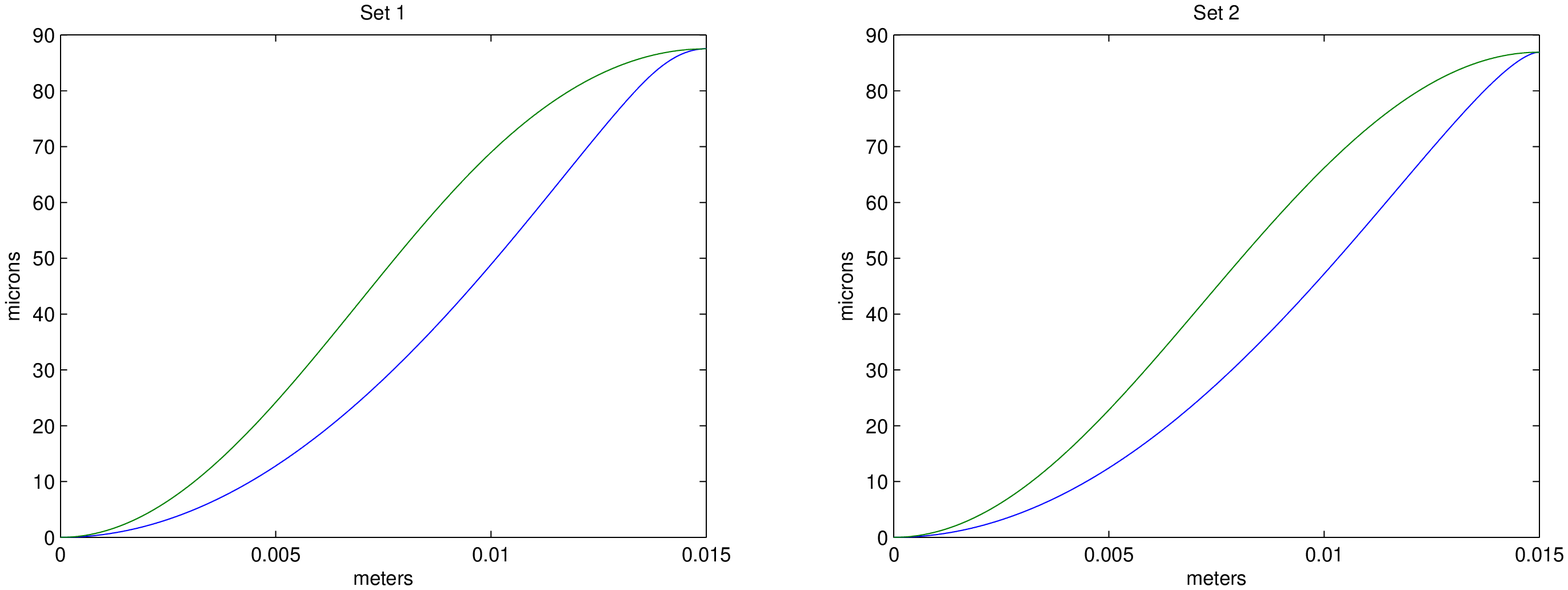}
\caption{The radial profiles of the four mirrors in the chained PIAA system.  The first mirror in each set is in blue, the second in green.}
\label{fig:4m}
\end{center}
\end{figure}

As expected, the system gains in inner working angle, to around $1.75 \lambda/D$, but its contrast is reduced by the additional diffraction in the system from two additional mirrors.  Both this plot and \myfig{fig:md1} are normalized to unity at the PSF peak; the throughput of the two-mirror system is $50.9\%$ on-axis, while the throughput of the four-mirror system is $87.9\%$.  In both cases, the first mirror pair has sag of $87.6\mu$m on each mirror; the second mirror pair has $86.9\mu$m of sag.  Both are shown in \myfig{fig:4m}  (For reference, both of these designs are on flat mirrors sized in accordance with the low-cost/low-sag design principles outlined in \mycitet{Bal10} and \mycitet{Bal11}.)  We note that this approach could also be identically applied to the analysis of a pair of deformable mirrors used in conjunction with a fixed set of PIAA mirrors, following the analysis in \mycitet{Sha07} and \mycitet{Pue11}.

\section{Summary}

In this work, we have presented three sets of equations which allow apodizations and mirror shapes to be designed exactly given choices for the remaining elements.  We expect in the near future to provide explicit representations for the residual diffraction in terms of the amplitude profiles presented here, and the derivation of these equations from an approximation to the diffraction integral provides a solid foundation for this future work.  (This approximation was also shown to reproduce the geometric-optics amplitude distribution for a uniform input field.)  A number of examples demonstrating the use of these equations have been provided; it is our hope that these will prove to be useful tools for subsequent PIAA designs.

We have also used these tools to design and analyze systems with multiple pairs of PIAA mirrors, to allow significantly-improved throughput as well as smaller inner working angle.  We expect to verify the performance of this class of system experimentally in the upcoming months; initial progress may be found in \mycitet{Bal11}.

\section{Acknowledgments}

The authors would like to thank L. Pueyo, S. Shaklan,  K. Balasubramanian, and T. Groff for ideas, feedback, and discussion over the course of this work, and L. Pueyo in particular for reading drafts of this manuscript.   The research was carried out at the Jet Propulsion Laboratory, California Institute of Technology, under a contract with the National Aeronautics and Space Administration.

\bibliography{refs}

\begin{thebibliography}{23}
\expandafter\ifx\csname natexlab\endcsname\relax\def\natexlab#1{#1}\fi

\bibitem[{Arfken(1984)}]{Arf84}
Arfken, G. 1984, Mathematical Methods for Physicists (Academic Press)

\bibitem[{{Balasubramanian} {et~al.}(2011){Balasubramanian}, {Cady}, {Pueyo},
  {An}, {Shaklan}, {Guyon}, \& {Belikov}}]{Bal11}
{Balasubramanian}, K., {Cady}, E., {Pueyo}, L., {et~al.} 2011, in Society of
  Photo-Optical Instrumentation Engineers (SPIE) Conference Series, Vol. 8151,
  Society of Photo-Optical Instrumentation Engineers (SPIE) Conference Series

\bibitem[{{Balasubramanian} {et~al.}(2010){Balasubramanian}, {Shaklan},
  {Pueyo}, {Wilson}, \& {Guyon}}]{Bal10}
{Balasubramanian}, K., {Shaklan}, S.~B., {Pueyo}, L., {Wilson}, D.~W., \&
  {Guyon}, O. 2010, in Society of Photo-Optical Instrumentation Engineers
  (SPIE) Conference Series, Vol. 7731, Society of Photo-Optical Instrumentation
  Engineers (SPIE) Conference Series

\bibitem[{{Belikov} {et~al.}(2006{\natexlab{a}}){Belikov}, {Give'on}, {Cady},
  {Kay}, {Pueyo}, \& {Kasdin}}]{Bel06}
{Belikov}, R., {Give'on}, A., {Cady}, E., {et~al.} 2006{\natexlab{a}}, in
  Bulletin of the American Astronomical Society, Vol.~38, Bulletin of the
  American Astronomical Society, 1131

\bibitem[{{Belikov} {et~al.}(2006{\natexlab{b}}){Belikov}, {Kasdin}, \&
  {Vanderbei}}]{Bel06a}
{Belikov}, R., {Kasdin}, N.~J., \& {Vanderbei}, R.~J. 2006{\natexlab{b}}, The
  Astrophysical Journal, 652, 833

\bibitem[{{Belikov} {et~al.}(2009){Belikov}, {Pluzhnik}, {Connelley},
  {Witteborn}, {Lynch}, {Cahoy}, {Guyon}, {Greene}, \& {McKelvey}}]{Bel09}
{Belikov}, R., {Pluzhnik}, E., {Connelley}, M.~S., {et~al.} 2009, in Presented
  at the Society of Photo-Optical Instrumentation Engineers (SPIE) Conference,
  Vol. 7440, Society of Photo-Optical Instrumentation Engineers (SPIE)
  Conference Series

\bibitem[{Born \& Wolf(1999)}]{Bor99}
Born, M., \& Wolf, E. 1999, Principles of Optics (Cambridge University Press)

\bibitem[{{Dickey} \& {Holswade}(2000)}]{Dic00}
{Dickey}, F.~M., \& {Holswade}, S.~C., eds. 2000, {Laser Beam Shaping: Theory
  and Techniques}

\bibitem[{Guyon(2003)}]{Guy03}
Guyon, O. 2003, Astronomy \& Astrophysics, 404, 379

\bibitem[{{Guyon} {et~al.}(2010){Guyon}, {Pluzhnik}, {Martinache}, {Totems},
  {Tanaka}, {Matsuo}, {Blain}, \& {Belikov}}]{Guy10}
{Guyon}, O., {Pluzhnik}, E., {Martinache}, F., {et~al.} 2010, Publications of
  the Astronomical Society of the Pacific, 122, 71

\bibitem[{{Guyon} {et~al.}(2005){Guyon}, {Pluzhnik}, {Galicher}, {Martinache},
  {Ridgway}, \& {Woodruff}}]{Guy05}
{Guyon}, O., {Pluzhnik}, E.~A., {Galicher}, R., {et~al.} 2005, The
  Astrophysical Journal, 622, 744

\bibitem[{{Kern} {et~al.}(2009){Kern}, {Belikov}, {Give'On}, {Guyon},
  {Kuhnert}, {Levine-West}, {McMichael}, {Moody}, {Niessner}, {Pueyo},
  {Shaklan}, {Traub}, \& {Trauger}}]{Ker09}
{Kern}, B., {Belikov}, R., {Give'On}, A., {et~al.} 2009, in Presented at the
  Society of Photo-Optical Instrumentation Engineers (SPIE) Conference, Vol.
  7440, Society of Photo-Optical Instrumentation Engineers (SPIE) Conference
  Series

\bibitem[{{Krist} {et~al.}(2010){Krist}, {Pueyo}, \& {Shaklan}}]{Kri10}
{Krist}, J.~E., {Pueyo}, L., \& {Shaklan}, S.~B. 2010, in Presented at the
  Society of Photo-Optical Instrumentation Engineers (SPIE) Conference, Vol.
  7731, Society of Photo-Optical Instrumentation Engineers (SPIE) Conference
  Series

\bibitem[{{Pluzhnik} {et~al.}(2006){Pluzhnik}, {Guyon}, {Ridgway},
  {Martinache}, {Woodruff}, {Blain}, \& {Galicher}}]{Plu06}
{Pluzhnik}, E.~A., {Guyon}, O., {Ridgway}, S.~T., {et~al.} 2006, The
  Astrophysical Journal, 644, 1246

\bibitem[{{Pueyo}(2008)}]{Pue08}
{Pueyo}, L. 2008, PhD thesis, Princeton University

\bibitem[{{Pueyo} {et~al.}(2011{\natexlab{a}}){Pueyo}, {Kasdin}, {Carlotti}, \&
  {Vanderbei}}]{Pue11a}
{Pueyo}, L., {Kasdin}, N.~J., {Carlotti}, A., \& {Vanderbei}, R.
  2011{\natexlab{a}}, The Astrophysical Journal Supplement Series, 195, 25

\bibitem[{{Pueyo} {et~al.}(2011{\natexlab{b}}){Pueyo}, {Kasdin}, \&
  {Shaklan}}]{Pue11}
{Pueyo}, L., {Kasdin}, N.~J., \& {Shaklan}, S. 2011{\natexlab{b}}, Journal of
  the Optical Society of America A, 28, 189

\bibitem[{{Pueyo} {et~al.}(2009){Pueyo}, {Shaklan}, {Give'On}, \&
  {Krist}}]{Pue09}
{Pueyo}, L., {Shaklan}, S., {Give'On}, A., \& {Krist}, J. 2009, in Presented at
  the Society of Photo-Optical Instrumentation Engineers (SPIE) Conference,
  Vol. 7440, Society of Photo-Optical Instrumentation Engineers (SPIE)
  Conference Series

\bibitem[{{Romero} \& {Dickey}(1996)}]{Rom96}
{Romero}, L.~A., \& {Dickey}, F.~M. 1996, Journal of the Optical Society of
  America A, 13, 751

\bibitem[{{Shaklan} {et~al.}(2007){Shaklan}, {Give'on}, {Belikov}, {Pueyo}, \&
  {Guyon}}]{Sha07}
{Shaklan}, S.~B., {Give'on}, A., {Belikov}, R., {Pueyo}, L., \& {Guyon}, O.
  2007, in Society of Photo-Optical Instrumentation Engineers (SPIE) Conference
  Series, Vol. 6693, Society of Photo-Optical Instrumentation Engineers (SPIE)
  Conference Series

\bibitem[{{Traub} \& {Vanderbei}(2003)}]{Tra03}
{Traub}, W.~A., \& {Vanderbei}, R.~J. 2003, The Astrophysical Journal, 599, 695

\bibitem[{Vanderbei(2006)}]{Van06}
Vanderbei, R. 2006, Astrophysical Journal, 636, 528

\bibitem[{{Vanderbei} \& {Traub}(2005)}]{Van05}
{Vanderbei}, R.~J., \& {Traub}, W.~A. 2005, The Astrophysical Journal, 626,
  1079

\end{thebibliography}
\bibliographystyle{apj}

\end{document}